\documentclass[english,english,english]{article}
\usepackage[T1]{fontenc}
\usepackage[latin1]{inputenc}
\usepackage{a4wide}
\usepackage{babel}
\makeatletter

\providecommand{\LyX}{L\kern-.1667em\lower.25em\hbox{Y}\kern-.125emX\@}

\usepackage[T1]{fontenc}
\usepackage[latin1]{inputenc}
\usepackage{a4wide}
\usepackage{babel}
\makeatletter

\usepackage[T1]{fontenc}
\usepackage[latin1]{inputenc}
\usepackage{a4wide}
\usepackage{babel}


\makeatother
\begin{document}
{\large \hfill{}}\textbf{\large Local indistinguishability and LOCC monotones \hfill{} }{\large \par}

\hfill{} Micha{\l} Horodecki\footnote{%
fizmh@univ.gda.pl 
}, Aditi Sen(De)\footnote{%
aditi@iftia6.univ.gda.pl 
}, Ujjwal Sen\footnote{%
ujjwal@univ.gda.pl 
} and Karol Horodecki\footnote{%
khorodec@manta.univ.gda.pl 
}\hfill{} 

{\footnotesize \hfill{}}\emph{\footnotesize Institute of
Theoretical Physics and Astrophysics, University of Gda\'{n}sk, 80-952
Gda\'{n}sk, Poland}{\footnotesize \hfill{}} {\footnotesize \par{}}{\footnotesize \par}

\begin{abstract}

We provide a method for checking indistinguishability of a set of 
multipartite orthogonal states by local operations and classical communication (LOCC). It 
bases on the principle of  nonincreasing of 
entanglement under LOCC.  This method originates from the one introduced 
by Ghosh \emph{et al.} (Phys. Rev. Lett. \textbf{87}, 5807 (2001) (quant-ph/0106148)), 
though we deal with {\it pure} states. 
In the bipartite case, our method 
is operational, although we do not know whether 
it can always detect local indistinguishability.
We apply our method to show that an arbitrary complete 
multipartite orthogonal basis 
is indistinguishable if it contains at least one entangled state. 
We also show that probabilistic distinguishing is possible for full basis
if and only if all vectors are product. We employ our method to 
prove local indistinguishability in a very interesting example
akin to "nonlocality without entanglement".

\end{abstract}

Orthogonal quantum state vectors can always be distinguished if there 
are no restrictions to measurements that one can perform. If the vectors  are 
states of a system consisting of two distant subsystems, then 
there can be  natural restrictions for the measurements that can be done. 
In particular, if Alice and Bob (the parties holding the subsystems) 
cannot communicate quantum information, their possibilities 
significantly decrease \cite{LOCC}.  Intuitively one feels that in such a case,
there will be a problem with distinguishing entangled states,
while product ones should remain distinguishable. The first result 
in this area was rather surprising: in Ref. \cite{nlwe} 
the authors exhibited a set of bipartite pure {\it product} states, that cannot be 
distinguished with certainty by local operations and classical communications (LOCC).
Another counterintuitive result was
obtained in Ref. \cite{Walgate}: {\it any} two orthogonal multipartite states
can be distinguished from each other by LOCC, irrespective of  
how entangled they are. The latter result was greatly 
extended in Refs. \cite{VSPM, CY}. There is therefore a general question: which sets 
of states are distinguishable?

To find that a given set {\it is} distinguishable, one usually needs 
to build suitable protocol. To show that the states are {\it not} 
distinguishable
one can try to eliminate all possible measurements as 
in \cite{Walgate2}. Another way  is to employ somehow the theory of 
entanglement 
\cite{huge, Plenio, Vidal-mon, miaryQIC}. A typical statement proving 
such indistinguishability would be 
then: Alice and Bob cannot distinguish the states, as they would   
increase entanglement otherwise (which is impossible by LOCC).
The advantage 
of the latter method is that it allows to estimate the entanglement 
resources  needed to distinguish the states, that are non-distinguishable 
by LOCC. 

In Ref. \cite{TDL-hiding} this approach was first used to check 
distinguishability between two mixed states (we will call it TDL method).  
Another powerful method  based on entanglement 
was recently  designed  in Ref. \cite{Aditi}  (we will call it 
GKRSS method). 
In this paper we introduce another method, closely related 
to the latter one, but connected also with the TDL method. 
Our approach
provides a strong tool for investigation of distinguishability of 
sets of bipartite pure states, because it bases on deciding whether some pure state
can be transformed into some other pure states by LOCC,  the latter 
issue  being completely solved in a series of papers on entanglement 
monotones and entanglement manipulations with pure states 
\cite{Vidal-mon, Nielsen, Vidal, JP}.
Using it, we show that any full basis of  an
arbitrary number of systems is not 
distinguishable, if  at least one  of vectors is entangled \cite{nieyeden}. For 
$2\otimes n$ system it is then also ``only if'', as product 
bases are distinguishable in this case \cite{UPB}. 
Our result applies also to probabilistic distinguishability, 
so that in conjunction with the result of \cite{Lewenstein} we obtain 
that a full basis is probabilistically distinguishable 
if and only if all vectors are product. 
As an illustration of the effectiveness 
of our presented method, 
we 
consider
some examples of local indistinguishability
of \emph{incomplete} bases. One of the examples exhibits 
an interesting feature akin to "nonlocality without entanglement" \cite{nlwe}.

Let us first note that the application of entanglement 
theory to this problem is not immediate. Imagine, that we want 
to distinguish between the four Bell states given by
\[
\begin{array}{rcl}
\displaystyle \left| B_{1}\right\rangle  & = & {\frac{1}{\sqrt{2}}\left( \left| 00\right\rangle +\left| 11\right\rangle \right) },\\
\left| B_{2}\right\rangle  & = & {\frac{1}{\sqrt{2}}\left( \left| 00\right\rangle -\left| 11\right\rangle \right) },\\
\left| B_{3}\right\rangle  & = & {\frac{1}{\sqrt{2}}\left( \left| 01\right\rangle +\left| 10\right\rangle \right) },\\
\left| B_{4}\right\rangle  & = & {\frac{1}{\sqrt{2}}\left( \left| 01\right\rangle -\left| 10\right\rangle \right) }.
\end{array}
\]
If we were able to apply  by LOCC just the von Neumann 
measurement, then we could obviously create entanglement. Namely,
if Alice and Bob start with any initial state (hence also possibly a disentangled one),
after  the von Neumann measurement, it collapses into one of Bell states.
This is of course impossible. We cannot however conclude at this 
moment,
that they are indistinguishable.  The clue is that we could 
distinguish between them, while destroying  them during the process. 
Thus Alice and Bob would get to know what state they shared, 
but the potential entanglement would be destroyed. This is 
actually the case in the Walgate \emph{et al.} protocol \cite{Walgate}, where 
one distinugishes between two (possibly) entangled states.

To employ entanglement theory in the distinguishability question, 
a more clever method should be applied. The general hint 
is to apply the measurement to some larger system. This concept is  
a basis for the TDL and GKRSS methods. In the first one 
\cite{TDL-hiding} the  authors  considered a state 
of four systems A, B, C, D:
\[
\psi=\psi_{AB}\otimes \psi_{CD}
\]
where $\psi_{AB}$ and $\psi_{CD}$ are maximally entangled states.
Then the measurement is applied to the AB part (cf. \cite{Lewenstein}). 
If the state 
after measurement is entangled, then one concludes that the measurement 
cannot be done  by use of LOCC.

The GKRSS method   \cite{Aditi} is the following. Given 
the set of states $\{\psi_i^{AB}\}_{i=1}^k$ to be distinguished, one 
builds a mixed state 
\begin{equation}
\varrho ={1\over k}\sum_{i} \left|\psi_i\right\rangle\left\langle\psi_i\right| \otimes 
\left|\phi_i\right\rangle\left\langle\phi_i\right|
\end{equation}
where $\phi_i$ are some entangled states of the CD system (more generally 
one could put some probabilities $p_i$ instead of $1/k$). Now
if Alice(A) and Bob(B)  are  able to distinguish between the states $\psi_i$ 
they can tell the result of their measurement to Claire(C) and Danny(D), who will 
then share states $\phi_i$ with probability $1/k$. One now compares 
the initial entanglement $E(\varrho)$ measured across the AC:BD cut 
and the final one  given by $(1/k)\sum_i E(\phi_i)$ according to any 
chosen entanglement measure $E$. 
If the states $\psi_i$ are  distinguishable by LOCC, then 
the final entanglement cannot be greater than the initial one; otherwise 
one could increase entanglement by LOCC \cite{monotone}. Thus, if 
we have 
\begin{equation}
\label{eq-aditi}
E(\varrho)< {1\over k}\sum_i E(\phi_i)
\end{equation}
then the states $\psi_i$ are not distinguishable by LOCC. 
In Refs. \cite{Aditi, Aditi2} distillable entanglement was used as 
$E$.

Let us now  exhibit the method of the present paper. It 
is a modification of the GKRSS method.  Namely instead 
of classical correlations between AB and CD we will use 
quantum correlations. Consequently mixture (\ref{eq-aditi})
is replaced by the {\it superposition}
\begin{equation}
\psi_{ABCD}=\sum_i \sqrt{p_i}\left|{\psi}_{i}^{AB}\right\rangle\left|{\phi}_{i}^{CD}\right\rangle
\label{pure}
\end{equation}
The states $\phi_i$ will be  used here 
essentially to \emph{detect} as to whether a set of states are locally 
indistinguishable and as 
such we shall henceforth call them "detectors".  At a first 
glance it seems that this approach should fail, because the pure state
is unlikely to have small entanglement. In \cite{Aditi} where 
mixtures are used, the possibility for the initial state $\varrho_{ABCD}$ 
to be separable in the AC:BD cut was much larger, as mixed states
are less coherent than pure ones; for a pure state to be separable,
it has to be product, while for mixed states, the very 
mixedness can decrease entanglement, or even produce 
separability \cite{volume}. Let us however exhibit 
the following example. Suppose that Alice and Bob are to distinguish between the Bell
 states  $\left|B_i\right\rangle$. As detectors, we take the same 
states (as in \cite{Aditi}). Our pure state is thus 
\begin{equation}
\label{Bell}
\left| \psi_{B} \right\rangle _{ABCD} =\frac{1}{2}\sum ^{4}_{i=1} \left| B_{i}\right\rangle _{AB}
\left| B_{i}\right\rangle _{CD} 
\end{equation}
One can see that this state can be written as 
\begin{equation}
\label{Michal}
\frac{1}{\sqrt{2}} \left(\left| 00\right\rangle + 
\left| 11\right\rangle\right)_{AC} \frac{1}{\sqrt{2}}
\left(\left| 00\right\rangle + \left| 11\right\rangle\right)_{BD} 
\end{equation}
So it turns out that it is \emph{product} in AC:BD cut, so that our 
method will work.  Assuming now the four Bell states to be locally 
distinguishable immediately would imply that 
the state \( \left| \psi \right\rangle \) is entangled in the AC:BD 
cut which is the desired contradiction. This result was obtained in 
\cite{Aditi} and their  mixed state 
$\varrho_{ABCD}={1\over 4}\sum_{i} \left| B_{i} \right\rangle \left\langle B_{i} \right| \otimes 
\left|B_{i} \right\rangle \left\langle B_{i} \right|$ 
turned out 
to be separable in AC:BD (see also \cite{smolin}). 
Here we have a pure state which is product. 
Note that in this particular example, our method, even though 
originating from the GKRSS approach,  coincides with the TDL method.

The advantage of our approach over the GKRSS method is that 
for mixed states, it is usually hard to check the relation (\ref{eq-aditi})
for different entanglement measures. Indeed,  for mixed states 
it is difficult to evaluate the known entanglement measures.
In our case we have pure states on both sides of the inequality, for which 
the set of all needed measures is known \cite{Vidal-mon, Vidal}.  
Even more:  Jonathan and Plenio \cite{JP},  generalizing the 
Nielsen result \cite{Nielsen},
have obtained a necessary and sufficient condition for the transformation 
from a pure state $\phi$ to an ensemble of pure states $\{p_i,\phi_i\}$.
The condition is efficiently computable. 
Namely, let $\lambda$ 
and $\lambda_i$  be vectors of the Schmidt coefficients
of $\phi$ and $\phi_i$ respectively.
Then the LOCC transition $\phi\to \{p_i,\phi_i\}$ is possible if
and only if the vector $\sum_ip_i\lambda_i$ majorizes $\lambda$
\cite{majorization}. To summarise, our method consists of 
the following steps
\begin{itemize}
\item[(1)] Given the states $\{\psi_i^{AB}\}_{i=1}^k$ to be distinguished, 
choose $k$ detectors $\phi_i^{CD}$ and probabilities $p_i$.  
\item[(2)]Applying the Jonathan-Plenio criterion, check if the transition 
$\psi_{ABCD}\to \{p_i,\phi_i^{CD}\}$ is possible by LOCC 
(in AC:BD cut) where  $\psi_{ABCD}$ is of the form  (\ref{pure}).
\end{itemize}

The item (1) can be formulated more generally in the following way:
\begin{itemize}
\item[(1a)] Choose $\psi_{ABCD}$ such that its reduction $\varrho_{AB}$ 
has the support spanned by $\psi_i^{AB}$'s. 
\item[(1b)]  Determine detectors $\phi_i^{CD}$ by writing 
$\psi_{ABCD}$ by means of $\psi_i^{AB}$. 
\end{itemize}

Now we will apply our method  to obtain the following proposition, 
where in fact we do not need an explicit use of the Jonathan-Plenio criterion.

{\bf Proposition.} Let $\psi_i^{AB}$ be a full orthogonal basis of an $m\otimes n$ 
system. Then we have: (1) If at least one of the vectors is entangled (see \cite{nieyeden}), 
the set cannot be perfectly distinguished by LOCC (2) The set 
cannot be probabilistically distinguished if and only if all vectors are product. 

{\bf Remark.} Note that we will not have "if and only if" for item (1)
because there are orthogonal product bases that cannot be distinguished 
\cite{nlwe}. However it would also be "only if" in \(2 \otimes n\),
as all product bases are locally distinguishable there \cite{UPB}.

{\bf Proof.} Consider the four party state  
\[
\left| \psi \right\rangle_{ABCD} = \left( \frac{1}{ \sqrt {m}}
\sum ^{m}_{i=1} \left| ii\right\rangle _{AC} \right) 
\left( \frac{1}{ \sqrt {n}}\sum ^{n}_{j=1} 
\left| jj\right\rangle _{BD} \right) 
\]
shared between Alice, Bob, Claire and Danny,
which is product across the AC:BD cut.
Note that Alice and Claire are sharing \( m \) -dimensional systems each while Bob and Danny are 
sharing \( n \) -dimensional systems each. Written in AB:CD, this state takes the form 
\begin{equation}
\label{bipartite}
\frac{1}{ \sqrt {mn}}\sum ^{mn}_{k=1} \left| k\right\rangle_{AB} \left| k\right\rangle_{CD} 
\end{equation}
However we know that such a state is \( U \otimes U^* \) invariant 
where \( U \) is an arbitrary
unitary operator on the mn-dimensional Hilbert space (with the tensor product
separating AB from CD) and 
where the complex conjugation is taken
in the computational basis (see e.g. \cite{xor}). We would choose 
our \( U \) as indicated below.

Let \( \left\{ \left| \psi_{1} \right\rangle, \left| \psi_{2} \right\rangle,..., \left| \psi_{mn} \right\rangle \right\} \) 
be a set of \( mn \) orthonormal states of an \( m \otimes n \) system. We choose our 
\( U \) such that \( U \left|k\right\rangle = \left|\psi_{k}\right\rangle \) for all 
\(k = 1, 2,..., mn \). We now use the \( U \otimes U^* \) invariance of the state 
\( \left|\psi\right\rangle \) in the AB:CD cut and write it as 
\[
\frac{1}{ \sqrt {mn}}\sum ^{mn}_{k=1} \left| \psi_{k}\right\rangle_{AB} \left| \psi_{k}\right\rangle^{*}_{CD} 
\]
where the complex conjugation is again in the computational basis.

Therefore if Alice and Bob are able to locally distinguish between the \( \left|\psi_{k}\right\rangle \)s,
they could ring up Claire and Danny to tell which state they share, resulting in the creation of
the corresponding correlated state \( \left|\psi_{k}\right\rangle^{*} \) 
between Claire and Danny.

Now if at least one among the \( \left|\psi_{k}\right\rangle \)s is entangled, an assumption
of local distinguishability of the \( \left|\psi_{k}\right\rangle \)s would imply that the 
state \( \left|\psi\right\rangle \) has a nonzero amount of entanglement in the 
AC:BD cut \cite{key-6}. But this is 
forbidden as \( \left|\psi\right\rangle \) is product in the AC:BD cut.

Note that the above reasoning goes through irrespective of whether the local distinguishability 
protocol for the \( \left|\psi_{k}\right\rangle \)s is deterministic or 
probabilistic. This 
proves that an arbitrary complete set of orthogonal states of any bipartite system is locally
indistinguishable (deterministically as well as probabilistically) if 
at least one of vectors is entangled. (Note that for the desired 
contradiction, the probabilistic protocol must have nonzero probability 
for at least one entangled state.) 
Now, from \cite{Lewenstein} it follows that any complete product basis
can be distinguished probabilistically \cite{Bennett}. Indeed in \cite{Lewenstein}
it was shown that any separable superoperator 
can be performed by LOCC with some probability of success. However,
measuring a complete product basis amounts to applying some 
separable superoperator. This ends the proof.









\textbf{Generalisation of the proposition.} 
The above proposition can in fact be generalised to the multiparty situation. 
That is, the following statement is true.
In \( d_{1} \otimes d_{2} \otimes \ldots  \otimes d_{N}\), 
(1) a full orthogonal basis
cannot be distinguished deterministically 
by LOCC (between the \(N\) sharing parties), if at least one of them is 
entangled (see \cite{nieyeden}) and
(2) a full orthogonal basis can be distinguished probabilistically 
if and only if all vectors are product (\emph{i.e.}, of the form
\(\left|\eta_1\right\rangle \otimes
 \left|\eta_2\right\rangle \otimes \ldots \otimes\left|\eta_N\right\rangle\)). 

Note that the entangled state that 
is needed for the validity of the statement of 
this generalised proposition may not be a genuine \(N\)-party entanglement.
For the \(2 \otimes 2 \otimes 2\) case, for example, even a state 
of the form \(\left(a\left|00\right\rangle + b\left|11\right\rangle\right)\otimes
\left|0\right\rangle\) in a complete orthogonal basis would be sufficient
for local indistinguishability of the basis. 

Item (1) of the generalised proposition is immediate,
from the Proposition for the bipartite case, once we 
note that a multiparty entangled state must 
be entangled in at least one bipartite cut. Note also
that if a set of multipartite states is indistinguishable 
in a bipartite cut, it would obviously remain so, if we lessen 
the allowed set of operations by restricting the parties 
within one cut to remain at distant locations. The "only if"
part of item (2) also follows by essentially the same argument. The "if" part
of item (2) however needs a generalisation of the result obtained in Ref. \cite{Lewenstein},
which we do now.

\def\<{\langle}

\def\>{\rangle}

Consider three systems $X,Y,Z$ with Hilbert spaces $C^d$ and fixed bases 
$\{|i\>\}$ (each of them will  be then considered as multipartite). 
Let $\varrho_Z$ be state we want to subject to 
operation $\Lambda$. Suppose also that systems $X$ and $Y$ 
are in the state $\sigma_{XY}=(\Lambda \otimes I_Y) P^+_{XY}$ where 
$P^+_{XY}=|\psi^+_{XY}\>\<\psi^+_{XY}|$, $\psi^+_{XY}=
{1\over \sqrt d}\sum_{i=1}^d |i\>_X|i\>_Y$. 
Following Ref. \cite{Lewenstein}  one finds that if the total state 
$\sigma_{XY}\otimes \varrho_Z$ is projected onto $I_X\otimes P^+_{YZ}$ 
(which happens with probability $1\over d^2$) then the state 
of the system $X$ becomes $\varrho^{out}_X=\Lambda(\varrho_Z)$.
Now, we take $X,Y,Z$ to be $n$-partite systems, so that e.g. 
$X=X_1X_2\ldots X_n$ with Hilbert space 
$C^{d_1}\otimes \ldots \otimes C^{d_n}$ and $d_1\ldots d_n=d$. The 
bases $\{|i\>\}$ are chosen to be product bases 
$|i_1\>\ldots |i_n\>$.  
If $\Lambda$ is separable operation (with respect 
to the considered division of $X$) then so is $\Lambda\otimes I_Y$,
as the set of separable operations is closed under tensor multiplication. 
The state $P^+_{XY}$ has the following product form 
$P^+_{XY}=P^+_{X_1Y_1}\otimes\ldots \otimes  P^+_{X_nY_n}$. 
Thus the state $\sigma_{XY}$ 
is separable. It then follows that $\Lambda$ can be performed 
by LOCC. To this end the parties prepare the state $\sigma_{XY}$
which is possible by LOCC, as it is separable.  Then $i$-th party  
perfom local projection into $P^+_{Y_iZ_i}$ 
which occurs with probability $1/d_i^2$. If all parties succeed,  
which happens  
with probability $1/d^2$, they obtain  projection $P^+_{YZ}$.

Let us 
illustrate the multiparty situation in a simple example in 
\(2\otimes2\otimes2\). Suppose that Alice, Bob and Claire want to distinguish between
the GHZ \cite{GHZ} states 
\[\begin{array}{rcl}
G_{1,2} & = & {\frac{1} {\sqrt 2} (\left | 000\right\rangle \pm \left | 111\right\rangle )},\\
G_{3, 4} & = & {\frac{1} {\sqrt 2} (\left | 010\right\rangle \pm \left | 101\right\rangle )},\\
G_{5, 6} & = & {\frac{1} {\sqrt 2} (\left | 100\right\rangle \pm \left | 011\right\rangle )},\\
G_{7, 8} & = & {\frac{1} {\sqrt 2} (\left | 001\right\rangle \pm \left | 110\right\rangle )}
\end{array}\]
Following the way we had proceeded for the 
four Bell states (in eqs. (\ref{Bell}) and (\ref{Michal})), we consider the pure state
\[
\left| \phi_{G} \right\rangle _{ABCDEF} =\frac{1}{2\sqrt{2}}\sum ^{8}_{i=1} \left| G_{i}\right\rangle _{ABC}
\left| G_{i}\right\rangle _{DEF} \]
This state can be rewritten as 
\[
\frac{1}{\sqrt{2}} \left(\left| 00\right\rangle + 
\left| 11\right\rangle\right)_{AD}  \frac{1}{\sqrt{2}}
\left(\left| 00\right\rangle + \left| 11\right\rangle\right)_{BE}\frac{1}{\sqrt{2}}
\left(\left| 00\right\rangle + \left| 11\right\rangle\right)_{CF}\]
That is, the state is a \emph{product}
 in  \(AD : BE : CF\). A 
similar argument as before, implies that the set of states \(\left\{G_{i}\right\}\) 
are indistinguishable 
by LOCC. 

Note however that our presented method for testing local 
indistinguishability of a set of \emph{bipartite} orthogonal states
cannot be extended in its full generality to the multipartite situation. 
The 
Jonathan-Plenio criterion \cite{JP} has not been as yet generalised to more 
than two parties. Given a set of tripartite (for definiteness)
orthogonal states \(\left\{\psi^{ABC}_{i}\right\}_{i=1}^{k}\) to be distinguished (locally),

(1) one may choose the \(k\) detectors \(\phi^{A_1 \ldots A_N}_{i}\) (\(N\) not 
necessarily equal to \(3\)) and probabilities \(p_i\) and

(2) see whether the transition \(\psi=\sum_{i}\sqrt{p_i}
\left|\psi^{ABC}_{i}\right\rangle\left|\phi^{A_1 \ldots A_N}_{i}\right\rangle
\rightarrow \left\{p_{i}, \phi^{A_1 \ldots A_N}_{i}\right\}\) is possible by a LOCC
protocol which is implementable by keeping \(A, B, C\) at distant locations.

If the transition is impossible, the set \(\left\{\psi^{ABC}_{i}\right\}_{i=1}^{k}\)
is indistinguishable when \(A, B, C\) are at distant locations. However 
in the absence of a criterion for transformation of (pure) states in 
the multipartite scenario, it would be in general hard to 
find out whether the above transformation in item (2) is possible or not \cite{bipartite}.
This is in contrast to the bipartite situation where 
our method is operational (via the Jonathan-Plenio criterion). Note 
however that this does \emph{not} imply that 
given a set of orthogonal bipartite states, our method would always 
detect its locally indistinguishability (if at all). 
We do not know whether our method does \emph{not} detect local indistinguishability
of some set of orthogonal states. We would mention later in this paper, as to why 
detecting local indistinguishability of a set of orthogonal states by our method
is interesting, even if we know it independently by other methods.

One can now try to see how effective the presented method is, when 
we deal with an \emph{incomplete} set of orthogonal states. In that direction,
we consider two examples. 

To discuss the first example, note that the set \(S\) consisting of the following 
maximally entangled states 
 in \(3 \otimes 3\) are distinguishable locally:
\begin{equation}
\label{max3x3}
\psi_{1}=\frac{1}{\sqrt{3}}\left(\left|00\right\rangle + \omega \left|11\right\rangle +  \omega^{2} \left|22\right\rangle\right),   
\psi_{2}=\frac{1}{\sqrt{3}}\left(\left|00\right\rangle + \omega^{2} \left|11\right\rangle + \omega \left|22\right\rangle\right),  
\psi_{3}=\frac{1}{\sqrt{3}}\left(\left|01\right\rangle + \left|12\right\rangle + \left|20\right\rangle\right)
\end{equation}
(\(\omega\) is a nonreal cube root of unity.) 
The set \(S\) can be distinguished locally by making a projective measurement 
in the basis 
\(\left\{\frac{1}{\sqrt{3}}\left(\left|0\right\rangle+\left|1\right\rangle+\left|2\right\rangle\right),
\frac{1}{\sqrt{3}}\left(\left|0\right\rangle+\omega\left|1\right\rangle+\omega^{2}\left|2\right\rangle\right),
\frac{1}{\sqrt{3}}\left(\left|0\right\rangle+\omega^{2}\left|1\right\rangle+\omega\left|2\right\rangle\right)\right\}\)
in any one of the parties and a subsequent classical communication to the 
other party (see also \cite{Dong}). 

Having shown this, what would be our expectation 
for the set of states containing the same states as in \(S\) but for the last state
\(\left|\psi_{3}\right\rangle\),
which is replaced by a \emph{product} state \(\left|01\right\rangle\)? 
The above Propositions seem to indicate that as we put more and more entanglement into 
the system, the system tends to become locally indistinguishable. This is also
the expectation obtained from the recent work of Walgate and Hardy \cite{Walgate2}. 
But one can check by taking \(B_{i}\)s \((i=1, 2, 3)\) as detectors and with probabilities
\(p_i\) as \(\left(.16, .16, .68\right)\), that 
the transition 
\(\sum_{i=1}^{3}\sqrt{p_{i}}\left|\psi_{i}\right\rangle_{AB}\left|B_{i}\right\rangle_{CD}
\rightarrow \left\{p_{i}, \left|B_i\right\rangle_{CD} \right\}\) is forbidden by the 
Jonathan-Plenio criterion \cite{JP}. 
Consequently the set \(S^{'}\), containing the states
\begin{equation}
\label{min3x3}
\psi_{1}=\frac{1}{\sqrt{3}}\left(\left|00\right\rangle + \omega \left|11\right\rangle +  \omega^{2} \left|22\right\rangle\right),   
\psi_{2}=\frac{1}{\sqrt{3}}\left(\left|00\right\rangle + \omega^{2} \left|11\right\rangle + \omega \left|22\right\rangle\right),  
\psi^{'}_{3}=\left|01\right\rangle
\end{equation}
are \emph{indistinguishable} by LOCC \cite{conjecture} \cite{block}.
This simple 
example shows that the intuition 
that we tried to obtain 
from our Propositions as well as from the work of 
Walgate and Hardy \cite{Walgate2} is not true.
In fact this example is more in the spirit
of the examples of "nonlocality without entanglement"
in Refs. \cite{nlwe, UPB, UPB1} as also the results in Refs. \cite{Walgate, VSPM, CY}
indicating that nonlocality (in the sense of local (in)distinguishability
of orthogonal multipartite states) is independent of entanglement.

We now go over to our second example of using our presented 
method to check indistinguishability of an incomplete basis. 
We would like to examine a case of indistinguishability of three orthogonal two-qubit states. 
This case has been solved in \cite{Walgate2}. But we want to solve it by our method. As we 
would see, it leads to an interesting open question.  
Consider again therefore a four party state   
\[
\left|\chi\right\rangle = \frac{1}{\sqrt{3}}\sum ^{3}_{i=1} \left| A_{i}\right\rangle _{AB}\left| B_{i}\right\rangle _{CD} \]
shared between Alice, Bob, Claire and Danny, 
to probe (by our method) the indistinguishability of the 
three orthogonal states \( \left| A_i \right\rangle \) given by
\[
\begin{array}{rcl}
\displaystyle \left| A_{1}\right\rangle  & = & a\left| 00\right\rangle + b\left| 11\right\rangle ,\\
\left| A_{2}\right\rangle  & = & b\left| 00\right\rangle - a\left| 11\right\rangle ,\\
\left| A_{3}\right\rangle  & = & {\frac{1}{\sqrt{2}}\left( \left| 01\right\rangle +\left| 10\right\rangle \right) },
\end{array}
\]
where \(a\) and \(b\) are real (with \(a^{2} + b^{2} = 1\)) 
and the detectors 
\( \left| B_i \right\rangle \) are 
the Bell states.  Let us again suppose that the three orthogonal 
states \(\left\{\left|A_{i}\right\rangle\right\}\) are 
distinguishable by LOCC
even if only a single copy is provided. But this implies that Alice and Bob would be able
to create the states \(\left| B_i \right\rangle \) \((i=1,2,3)\) (each with probability 1/3) 
between Claire and Danny.
Thus from the state \( \left| \chi  \right\rangle \), in the AC:BD cut,
it would be possible to create the states \(\left| B_i \right\rangle \) (i=1,2,3), each with 
probability 1/3, by LOCC only.  
According to  the  Jonathan-Plenio result \cite{orNielsen} the process 
is impossible, if one of the squares of the Schmidt coefficients of 
$\psi_{ABCD}$  across  AC:BD cut is smaller than $1/2$.
 We see 
that this is  the case when \(a\) satisfies \(.0252632 < a < .999681\). 
Thus the \(\left| A_i \right\rangle \)s \((i=1,2,3)\) are locally 
indistinguishable whenever \(a\) falls in the above range.   However 
as shown recently by  Walgate and Hardy \cite{Walgate2} three 
two-qubit vectors can be distinguished if only one of them is 
entangled \cite{Aditi2}.
Thus we should be able to show that the process of 
distinguishing  by LOCC is impossible   within the whole 
range $0 < a < 1$. We have tried with many different 
detectors, but the Bell states are most probably the optimal one. 
This intuition comes from the feeling that maximally entangled states 
would be the hardest to create.  
Therefore, change of detectors would possibly not produce the desired impossibility.
We can however achieve  it by  putting probabilities:
\[
p_1=p_2=1/4, \quad p_3=1/2
\]
instead of $p_i=1/3$.  For such probabilities we obtain
that distinguishing between the states \(\left| A_i \right\rangle \) leads to 
increasing some entanglement monotone in  the whole range of 
parameter $a$. This is the reason why we wanted to prove 
local indistinguishibility of the these states by our method even
when the result itself is known by other methods. Because 
proving local indistinguishability 
(of a set of orthogonal states) through our method immediately 
shows that any (nonlocal) superoperator which distinguishes
between these states would necessarily increase 
some LOCC monotone.

Since our method is based on entanglement monotones, there is a question, whether all 
operations that  cannot be performed by LOCC would increase at least 
one function monotonic under LOCC. Most likely it is the case, 
i.e. the set of LOCC doable operations is described by the 
set of  LOCC monotones.

There are tasks that can be implemented by separable superoperators,  but 
not by LOCC ones. An example is to distinguish the basis 
consisting of  product states given in Ref. \cite{nlwe}. An interesting question 
is whether our method can prove this indistinguishability. To this end, one would need 
a monotone, that {\it is not }  monotonic under separable operations \cite{rains}. 
In our method we go from pure states to pure states, and the set 
of monotones that are responsible for such possibility is finite \cite{Vidal-mon,JP}.
They are sums of squares of $k$ largest Schmidt coefficients ($k=1,\ldots,d$ where $d$ is 
dimension of subsystem).  There remains 
an open question whether they could increase under separable superoperators.
If the answer is ``yes'', then our  method might work also for
distinguishing product basis.  It is however clear that we could not then 
apply our methos with the initial state as product w.r.t $AC:BD$ cut. Indeed, separable superoperators 
cannot produce entangled state out of product ones, but can distinguish 
between the states of interest.

There is example of set of product vectors indistinguishable by 
LOCC, which is not full basis -- so called unextendible product basis (UPB) \cite{UPB}.
In the case of two qutrits it is known \cite{UPB1} that UPB can also be 
distinguished  by separable operations. 
However, for higher dimensions it might happen that even PPT operations
cannot distinguish an UPB, so that the conditions for monotones  would be less stringent.
Indeed, then  LOCC  monotones could be monotonic also under separable and PPT operations,
and could nevertheless be still useful.

We are grateful to Charles Bennett for drawing our attention to the 
fact, that a complete set of orthogonal product states are always probabilistically 
distinguishable locally at the European 
Research 
Conference on 
Quantum Information in San Feliu de Guixols, Spain, March, 2002.
We would like to thank Sibasish Ghosh, Guruprasad Kar, Anirban Roy, 
Debasis Sarkar and Barbara Synak for helpful discussions. 
The work is supported by the European Community under project EQUIP,
Contract No. IST-11053-1999 and by the University of Gda\'{n}sk, Grant No. BW/5400-5-0236-2.


\begin{thebibliography}{10}
\bibitem{LOCC}The class of operations thay can perform 
is then called LOCC (local operations and classical
communication) \cite{huge}.
\bibitem{monotone} In the Vidal approach \cite{Vidal-mon}, entanglement is defined simply
as a quantity that does not increase under LOCC (cf. \cite{RP}).
\bibitem{majorization}
Suppose $x=(x_1,\ldots, x_d)$ and  $y=(y_1,\ldots, y_d)$ are real 
$d$ dimensional vectors.
Then $x \prec y$ ($x$ is majorized by $y$) if for each $k$ in the 
range $1,\ldots, d$  
\[
\sum_j^k x_{j}^{\downarrow} \leq \sum_j^k y_{j}^{\downarrow}
\]
where \(x_{j}^{\downarrow}\) and \(y_{j}^{\downarrow}\) are 
elements of the vectors \(x\) and \(y\) respectively with their entries set in
decreasing order.
\bibitem{key-6} Note that \(\left|\phi\right\rangle^{*}\) is 
entangled whenever 
\(\left|\phi\right\rangle\) is entangled.
\bibitem{bipartite} One should first check the indistinguishability
of the states in all bipartite cuts. If a set of states 
is already indistinguishable in some bipartite cut, it would remain so 
with parties within the cut separated. It may also
be useful to consider the multiparty case (here tripartite), when 
nothing conclusive is obtained by considering the bipartite cuts.
\bibitem{conjecture} The result that we obtained with the three Bell state detectors
and with the displayed probabilities was not only for the state \(\left|01\right\rangle\).
Fixing the states \(\psi_1\) and \(\psi_2\), if we choose the three Bell detectors
and the probabilities \((.16, .16, .68)\), the three states 
\(\psi_1\), \(\psi_2\) and \(a\left|01\right\rangle + b\left|12\right\rangle + 
c\left|20\right\rangle\) are locally indistinguishable for a continuous region 
around \((a, b, c) = (1, 0, 0)\). We are willing to conjecture
that \(\psi_1\), \(\psi_2\) and \(a\left|01\right\rangle + b\left|12\right\rangle + 
c\left|20\right\rangle\) are locally indistinguishable 
for the whole range of \((a, b, c)\) except when the last is maximally entangled.
\bibitem{block} The example was considered
by keeping in mind that it may be interesting (with
respect to the local indistinguishability problem) to divide 
the Hilbert space into a block structure. For simplicity, let us
consider the \(3\otimes 3\) case only. The \(3 \otimes 3\) Hilbert space
can be thought of as divided into the three blocks (subspaces) spanned by 
\(\left\{\left|00\right\rangle, \left|11\right\rangle, \left|22\right\rangle\right\}\),
\(\left\{\left|01\right\rangle, \left|12\right\rangle, \left|20\right\rangle\right\}\) and 
\(\left\{\left|10\right\rangle, \left|21\right\rangle, \left|02\right\rangle\right\}\).
We chose two maximally entangled states from the first block and the product 
state from the second block. But any other combination 
of blocks could also have been used to show local indistinguishability.
\bibitem{orNielsen} Actually in the considered case, all final states 
have the same $\lambda_i$, so that the Jonathan-Plenio criterion 
reduces to Nielsen's one \cite{Nielsen}.
\end{thebibliography}
\end{document}